# Mesh-based segmentation for automated margin line generation on incisors receiving crown treatment

**Ammar Alsheghri** [1,5] *****, **Ying Zhang** [2], **Farnoosh Ghadiri** [3], **Julia Keren** [4], **Farida Cheriet** [2] and **Francois Guibault** [2]

[1]  Mechanical Engineering Department, King Fahd University of Petroleum and Minerals (KFUPM), Dhahran, 31261, KSA
[2]  Department of Computer and Software Engineering, Polytechnique Montréal, Montréal, QC, Canada
[3]  Centre d'intelligence artificielle appliquée (JACOBB), Montréal, QC, Canada
[4]  Intellident Dentaire Inc., Montréal, QC, Canada
[5]  Interdisciplinary research center for Biosystems and Machines, King Fahd University of Petroleum and Minerals (KFUPM), Dhahran, 31261, KSA

*****  Correspondence: ammar.sheghri@kfupm.edu.sa

**Abstract:** Dental crowns are essential dental treatments for restoring damaged or missing teeth of patients. Recent design approaches of dental crowns are carried out using commercial dental design software. Once a scan of a preparation is uploaded to the software, a dental technician needs to manually define a precise margin line on the preparation surface which constitutes a nonrepeatable and inconsistent procedure. This work proposes a new framework to determine margin lines automatically and accurately using deep learning. A dataset of incisor teeth was provided by a collaborating dental laboratory to train a deep learning segmentation model. A mesh-based neural network was modified by changing its input channels and used to segment the prepared tooth in two regions such that the margin line is contained within the boundary faces separating the two regions. Next, k-fold cross-validation was used to train 5 models and a voting classifier technique was used to combine their results to enhance the segmentation. After that, boundary smoothening and optimization using the graph cut method was applied to refine the segmentation results. Then, boundary faces separating the two regions were selected to represent the margin line faces. A spline was approximated to best fit the centers of the boundary faces to predict the margin line. Our results show that an ensemble model combined with maximum probability predicted the highest number of successful test cases (7 out of 13) based on a maximum distance threshold of 200 μm (representing human error) between the predicted and ground truth point clouds. It was also demonstrated that the better the quality of the preparation, the smaller the divergence between the predicted and ground truth margin lines (Spearman's rank correlation coefficient of -0.683). We provide the train and test datasets for the community[1].

**Keywords:** Tooth restoration design; Mesh-based segmentation; 3D Deep learning; Digital dentistry, Dental preparation margin line. Artificial intelligence; Digital systems

## 1. Introduction

The extraction of a margin line from a dental preparation is the first and most important step in the crown generation process. A dental crown is a treatment used to restore a patient's damaged tooth and it is considered among the most essential components of dental restorations. To receive a crown treatment, a dental preparation is created from the damaged tooth by removing the diseased tooth sections and molding the patient's tooth. Figure 1 shows a typical crown restoration procedure where a designed crown is placed on a dental preparation. The dental crown sits on the prepared tooth and seals it at the finish line or the margin line as shown in Figure 1. The margin line is defined as the terminal/peripheral portion of the tooth preparation [1]. This structure is essential for adequate seating of the crown [2].

---

[1] https://github.com/intellident-ai/teethPreparationData/tree/main.

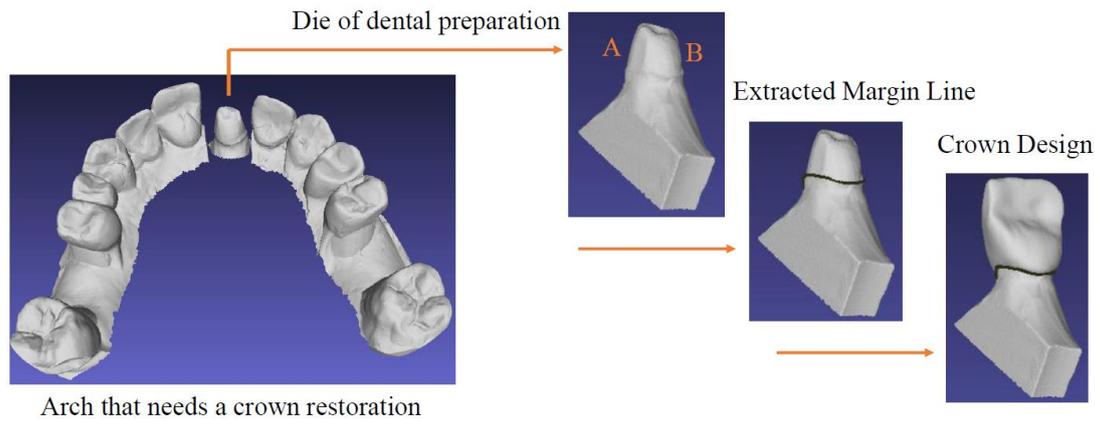

**Figure 1.** The accurate extraction of the margin line is essential for a successful crown generation. Accurate extraction of the margin line is challenging in areas of a dental preparation that are flat such as the surface A (i.e., fuzzy region that has no geometric features) whereas it is easier for surfaces of high curvature such as surface B.

While preparing a tooth to receive a dental crown, the dentist locates an appropriate margin line on the damaged tooth such that undercuts are avoided. It is essential to prepare the tooth with an adequate margin line for the success of the dental crown treatment [3, 4]. In practice, dentists try to conserve the maximum amount of tooth structure while preparing a tooth to create enough space and ensure the placement of a retentive and esthetic crown. The preparation should help the retention by resisting the crown restoration from removal along its path of insertion. Once a die is received by a dental technician to create a crown, a commercial dental restoration software is used by the technician to manually select the points of the margin line on the die. The margin line extraction process is a critical and difficult step in the crown restoration procedure. Despite advances in digital dentistry, the manual selection process makes the procedure non-repeatable, variable, and time-consuming depending on the experience of the dental technician [5]. Changes in the margin line design can result in differences in marginal adaptation of crowns [6] and can also affect the failure load and fractography of the crown [7]. It was reported that around 12.7% of dental crown restorations suffer from poor retention due to food residue coming from poor fitting at the gingival margin [8].

Extraction of margin lines from a 3D die geometry has been previously investigated in the literature [9]. Nevertheless, the problem of accurately determining a margin line is difficult because the dental margin line does not necessarily always follow the feature line as shown in Figure 1. There is very few published research on methods of margin line extraction. In [10] a margin line extracting algorithm based on a heuristic searching strategy based on feature information was presented. In [11], the A* algorithm was used to search for the minimum cost path for a graph that represents the outermost feature curve. However, such techniques still depend on the presence of a feature line region which could be missing in certain regions.

AI models are being increasingly and successfully used in digital dentistry including treatment planning in prosthodontics, orthodontics, and esthetic dentistry procedures [12-13]. In [14] deep learning was used to generate a crown shell for a dental restoration using multi-resolution generative neural networks. Later, transformers were used to replace neural networks in an attempt to improve the generation quality [15]. A detailed end-to-end framework for dental crown design using AI with preprocessing and postprocessing procedures are presented in [16]. In [17] a deep adversarial network-driven gingival margin line reconstruction framework was introduced to automatically obtain the personalized gingival contour for a partially edentulous patient. This work achieved superior performance compared to recent advances on real-world dental databases by preserving gingival contour details, structure, and perceptual features through a dual generator model and two-scale discriminator model, and reconstructing missing gingival margin lines harmoniously with adjacent teeth. Nevertheless, margin line reconstruction on surfaces of prepared teeth for dental crown treatment was not discussed. The features distinguishing margin lines between teeth and gingiva are different than the those distinguishing margin lines on surfaces of prepared teeth. Other studies used deep learning for teeth segmentation and labeling [18-20]. In the context of 3D surface data coming from optical scanners in the form of a triangular mesh, segmentation refers to distinguishing triangles of different classes or groups with different labels. Regarding automated 3D teeth segmentation from intraoral scans and labeling, researchers have achieved high efficiency using deep learning for applications in orthodontic diagnosis and appliance fabrication [21]. On the other hand, only a single study focused on the use of deep learning for margin line extraction from surfaces of prepared teeth with a dataset of 380 cases and reported enhanced results of automatically generating a margin line using AI compared with traditional methods [8, 22]. The S-Octree technique was used to divide the dental preparation into two parts along the feature line representing the margin line and providing ground truth labels to train a segmentation neural network. However, this might not be suited to provide ground truth data especially for dies with an abundance of fuzzy regions that have missing feature lines as shown, for example, by surface 'A' in Figure 1.

In practice, a dental crown is hollowed from inside and it consists of two main parts: a crown bottom that approximately has the same topology as the dental preparation region above the margin line, and a shell that provides the external 3D shape of a tooth (see Figure 2). Given the available ground truth data of a crown bottom, our approach focuses on segmentation of the region supporting the crown bottom component from the rest of the die using deep learning, then extracting the margin line from the boundary of the segmented region. The use of crowns designed by dental technicians to map ground truth labels on dies is an apparent contribution of our work compared with [8], which used an S-Octree model to provide ground truth labeling on the dies. Moreover, only four test cases were qualitatively presented in [8] out of forty test cases and no quantitative statistics were reported for results in that study.

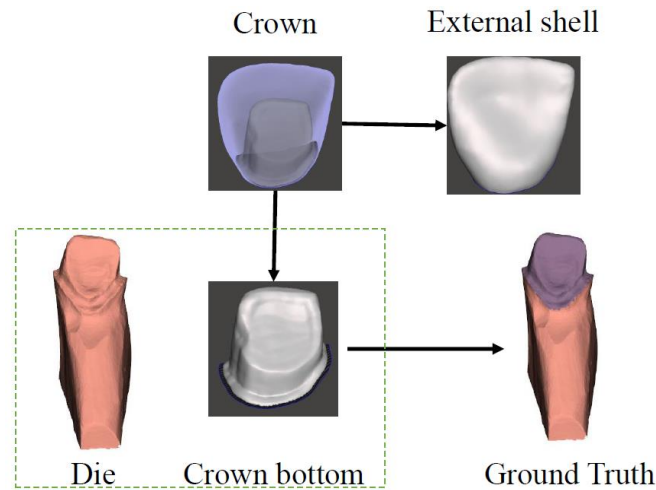

**Figure 2.** Labeling ground truth die. The crown-bottom part of the crown was extracted from the crown as shown and its boundary-band inner curve was used with the die scan to label the region above the margin line.

Because dental preparations are characterized with variable geometries that differ according to the tooth position and patient, existing methods still cannot meet the requirements for both efficiency and accuracy. Hence, an accurate and automatic extraction of the margin line technique is still missing, and it is the main objective of this manuscript to address this challenge. The methods are presented in the next section, followed by results, discussion, and conclusion.

## 2. Materials and Methods

### 2.1. Data preparation and labeling

In this study we used 54 die cases of oral patient's dental preparation models including incisors from the upper and lower arches. The standard labels of the selected incisors are 11, 21, 31, and 41. Out of all data, 41 samples (almost 75%) were used for training and 13 samples (almost 25%) were used for testing. According to dental technicians, identifying margin lines on incisors is more difficult than molars or premolars due to the possible absence of features. Therefore, this study focused on incisors only.

The input data consists of 3D surface scans of dies along with their corresponding master arches and dental crowns in STL mesh format. Crowns designed by dental professionals were used to extract the ground truth margin lines. The ground truth margin lines were obtained by extracting the boundary from the crown bottom by selecting edges which belong to single triangles. Margin line points were then obtained from the vertices of the boundary edges. The margin line points were defined as the vertices that lie on the surface of the crown bottom that intersects the die mesh. Next, the closest triangles to the margin line points were selected from the die mesh, which represents margin line faces. The margin line faces split the die mesh into two regions with two distinct labels: the upper region represents the crown bottom that contains the margin line faces as its boundary (purple region in Figure 2) while the bottom region represents the rest of the die (Figure 2). As the margin line triangular faces become smaller, their centroids provide a better approximation of the ground truth margin lines.

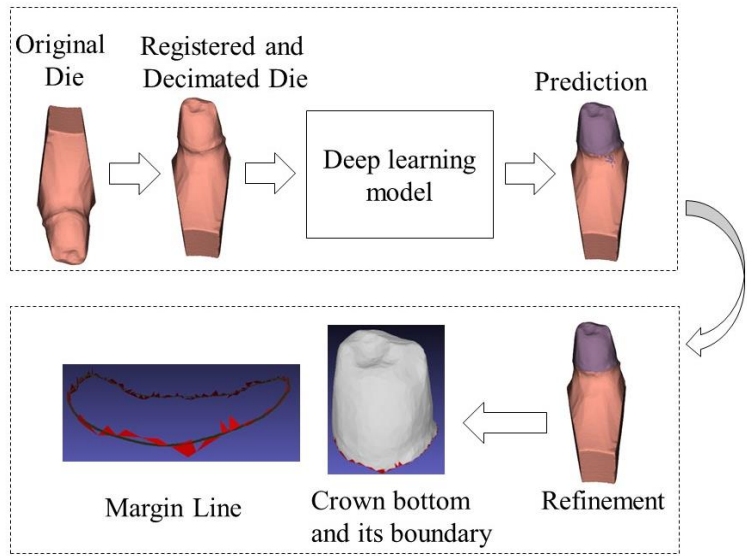

**Figure 3.** Overall inference framework based on a pre-trained deep learning model to segment prepped teeth. The input die is registered and decimated to 10 k triangles. The die is next segmented using the pre-trained segmentation model into two regions. The segmentation results are refined using the graph-cut technique which applies boundary smoothing. The boundary triangles of the upper region (i.e. crown bottom) are extracted and their centers are used to interpolate a spline representing the generated margin line.

### 2.2. Data preprocessing

Data preprocessing involved rigid registration and decimation. Labeled dies belonging to upper arches were rotated 180° such that they share the same orientation with dies belonging to lower arches (see Figure A1 in Appendix). This was done through oriented bounding box registration which uses a convex hull and then applies principal component analysis. After that, dies were decimated to reduce the number of triangles in the mesh to 10,000 triangles.

Augmentation was applied to increase the amount of data such that 20 samples were generated from each die by random rotations in the ranges of [-45°, 45°] around the x and y axes, [-180°, 180°] around the z axis, and scaling in the range of [0.9, 1.1] in the x, y, and z dimensions. The selected ranges were chosen such that the oriented bounding box registration is not excessively disrupted. For instance, rotation of the samples around the x and y axes was limited as opposed to the z axis. Finally, dies were normalized by subtracting the mean and dividing by standard deviation of their coordinates. The final number of training data after augmentation was 861 samples.

### 2.3. Segmentation deep learning network

The state-of-the-art MeshSegNet architecture [20] was chosen as the segmentation network for supervised training which is based on pointNet++ [23] and includes curvature features to improve teeth segmentation. Segmentation in this setting means to assign a label to every triangle on the die mesh to classify whether it belongs to the region of the preparation which supports the crown bottom or not. MeshSegNet works on various raw surface attributes as inputs including coordinates and normal vectors. It extracts multi-scale local contextual features hierarchically by integrating a sequence of graph-constrained learning modules, each fed by a multi-layer perceptron (MLP) along its forward pass. Next, local-to-global geometric characteristics are combined using a dense fusion technique to learn higher-level features for mesh cell labelling [20]. The output of the dense fusion block is passed to another MLP before applying a 1D convolution to obtain class probabilities. Figure 4 shows the network architecture [20]. This network was preferred to other semantic segmentation models [19] due to its proven ability to distinguish between teeth and gingiva. We modified the network by including 3 more input channels to account for curvature.

The training data was split into 5 folds. Four folds were used during training and a single fold was used as a validation set. The validation fold was shifted which resulted in 5 different trained models. An ensembled approach was used by combining results from the five different models. The model hyperparameters were selected as follows: the learning rate was $1 \times 10^{-3}$, the train and validation batch sizes were 10, Adam was used as the optimizer, the patch size was 10000, and the number of used classes was 2. The number of input channels was varied between 9 and 18 input channels to investigate the effect of different input channels on the model performance. The input channels entailed 9 channels for vertex cartesian coordinates, 3 channels for the cartesian coordinates of the barycenter of each triangle in the mesh, 3 channels for the normal vectors of the triangles' centers, and 3 channels for the discrete mean curvatures of each triangle's vertices. The discrete mean curvatures were calculated by selecting a characteristic radius for every mesh that equals the maximum edge length of that mesh. All other model parameters were kept as default based on the architecture of MeshSegNet [20].

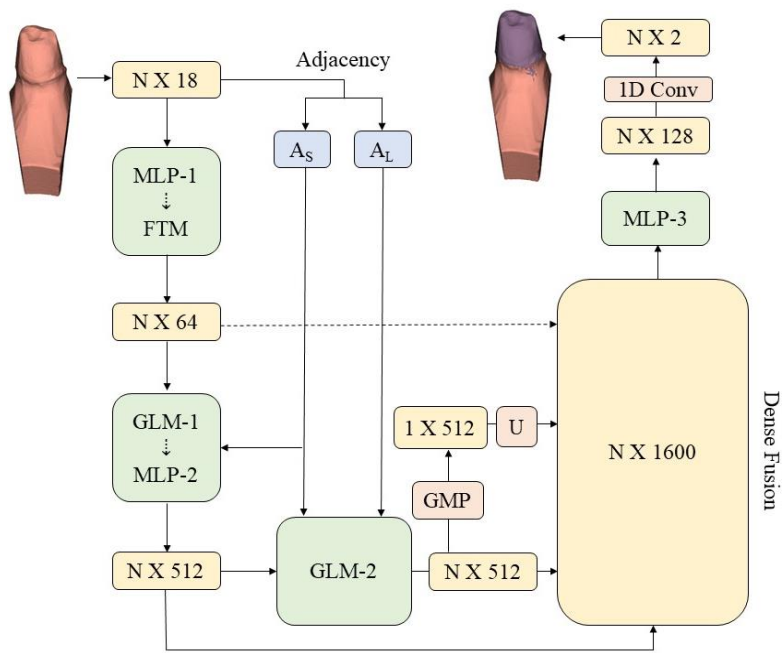

**Figure 4**: Architecture of MeshSegNet with modified input and output channels. MLP: Multi-layer perceptron; FTM: Feature-transformer module; GLM: graph constrained learning module with symmetric average pooling; GMP: Global matrix pooling; U: Upsampling; AS: Small scale adjacency matrix; AL: Large scale adjacency matrix.

The training was carried out for 200 epochs/iterations for all considered models. The training loss was used to update the weights of the model whereas the validation loss was used to choose the best model. Four segmentation models with different configurations were compared to evaluate the effect of different model parameters on the segmentation performance. Table 1 lists the different models with the main differences in the configurations.

*2.4. Postprocessing and margin line extraction*

Once the crown bottom region was predicted on the decimated die, geometric post processing for boundary optimization was applied using the graph-cut method to refine the prediction. This was done by combining the graph cut method and the labelling probabilities to refine the segmentation results and enhance the smoothening of the boundaries [24, 25]. After that, the die was re-oriented to the original position. To obtain a margin line, the region corresponding to the crown bottom on the preparation was separated from the rest of the die and its boundary faces were selected to represent the margin line faces. The boundary face centers were obtained and used to find the closest points on the surface of the original die scan. A B-spline was then interpolated through that set of 3D points with a trade-off between closeness and smoothness [33]. The scipy.interpolate.splprep method was used for B-spline fitting with a smoothness value of $0.005(N - \sqrt{2N})$, where N is the number of data points. Finally, the points on the surface of the die that are closest to the interpolated spline were located and chosen to represent the points of the margin line.

**Table 1**
Models' configurations used for different experiments

| Model number | Inclusion of points coordinates channels | Inclusion of curvature channels | Count of cells after decimation |
|---|---|---|---|
| 1 | Yes | Yes | 10,000 |
| 2 | Yes | No | 10,000 |
| 3 | Yes | Yes | 20,000 |
| 4 | No | Yes | 10,000 |

### 2.5. Ensemble models

We used a cross-validation technique and a voting classifier to combine the results coming from the 5-folds of the trained MeshSegNet deep learning model. Two techniques were investigated for the voting classifier: maximum probability and democracy. In maximum probability, the model that returned the maximum probability for a label was selected to assign that label. In democracy, a voting strategy among the five folds was considered to select the appropriate label of a triangle. Both maximum probability and democracy techniques were compared with the five different folds of the best model selected from Table 1.

### 2.6. Overall Framework of the proposed method

The proposed framework entails two major components: 1) Training a deep learning segmentation model to segment dies or prepped teeth; 2) Using the deep learning model to segment dies at inference stage then extracting the margin line. In both steps, the input die data should be registered and decimated to only 10 k triangles. During the training step, the input dies were labeled as described in sections 2.1 (data preparation and labeling) and augmented as described in section 2.2 (data preprocessing). Figure 3 illustrates the inference stage where the input data consists of unlabeled dies. After registration, the dies are segmented using the pre-trained model and then post-processed. The post-processing process involves refining the segmentation results using the graph-cut method and then extracting the margin line as described in sec 2.4 (Postprocessing and margin line extraction).

### 2.7. Validation methods

### 2.7.1. Metrics

Three metrics were used to quantitatively evaluate the performance of the segmentation model. Namely, dice similarity coefficient (DSC), positive predictive value (PPV), and sensitivity (SEN) whose values range between 0 and 1 and are positively proportional to the segmentation performance. The sensitivity, also called recall or true positive rate, measures the portion of positive triangles in the ground truth that are also identified as positive by the segmentation being evaluated [26]. The positive predictive value or the precision is the ratio of relevant true positive predictions among the positive predictions. Dice similarity coefficient is also called the overlap index or the F1 score, is the most used metric in direct comparison between automatic and ground truth segmentations [26]. The DSC is a harmonic average of precision and recall and is used to measure the accuracy of the test. The prediction metrics were calculated in Eqs. (1-3) based on the predicted labels of the mesh triangles being true positive (TP), false positive (FP), true negative (TN), and false negative (FN).

$$DSC = F_1\ score = accuracy = \frac{2TP}{2TP + FP + FN} \qquad (1)$$

$$SEN = Sensitivity = Recall = True\ positive\ rate = \frac{TP}{TP + FN} \quad (2)$$

$$PPV = precision = \frac{TP}{TP + FP} \qquad (3)$$

For more details, a qualitative illustration is provided in Figure 5. Two additional metrics were also reported to evaluate the predicted margin line which are the maximum and mean distances between the predicted and true margin lines. These distances were calculated based on the minimum spatial Euclidian distance between the two-point clouds of the predicted and ground truth margin lines. Finally, we reported the overall number of successful test cases based on a threshold defined by the acceptable human error produced by a dental technician while manually creating margin lines. The value of the threshold was 200 μm. More details are presented in the Results and Discussion sections.

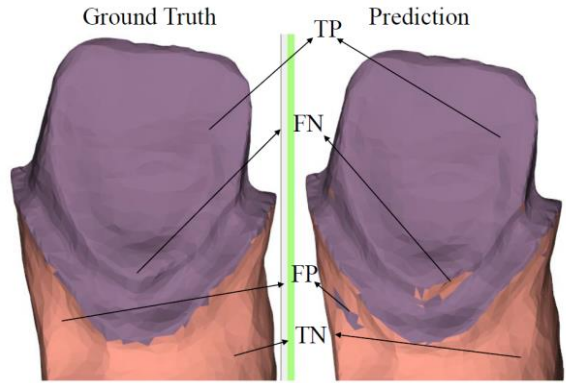

**Figure 5.** Qualitative illustration of predicted labels; TP: true positive, FP: false positive, TN: true negative, and FN: false negative.

### 2.7.2. Rating of preps and margin lines

The quality of ground truth margin lines was rated by a collaborating dental technician. Due to the absence of standardized criteria for the evaluation of preparations, we depended on guidelines from published literature [1, 27-28] as well as the opinion of an expert dental lab technician to rate the quality of the preparation (Kerenor Dental Studio, Montreal, Canada). For the score, we considered the margin line (visibility, regularity, smoothness, single or double margin line), the space for the crown thickness, the existence of sharp feature edges on the top of the prepared tooth, and the presence of undercuts. The rating was scored out of 4 such that 4 is very good, 3 is good, 2 is acceptable, 1 unworkable. The rating procedure was iterative such that two dental technicians qualitatively evaluated the preparations against the guidelines available in the literature [27-29].

### 2.7.3. Statistics

We used the Shapiro-Wilk test to check normality of the data. This test showed that the values of the segmentation metrics reported in Table 2 and Table 3 are not normally distributed. Therefore, the non-parametric Kruskal-Wallis test was used to investigate the significance among the values of the segmentation metrics reported in Table 2 and Table 3. Bonferroni's test was used for post hoc multiple comparisons. All statistical analyses were performed using R software, and p values of less than 0.05 were considered as indicating statistical significance.

## 3. Results

The quantitative results comparing the different models are presented in Table 2 and Table 3. While there was no statistical significance among the ranks of four model configurations in Table 2 in terms of the segmentation metrics, Model 1 scored the best in terms of the number of successfully predicted test cases. Therefore, the configurations of model 1 (i.e. inclusion of coordinates and curvature channels) were used to study the effect of using an ensemble technique which combined results from the five folds of model 1.

### 3.1. Quantitative and qualitative results of best model

All predicted margin lines seal the preparation with zero distance from the preparation 3D surface. Table 3 shows the results of the five folds and the combined models. The best overall model outcomes were for the combined models with maximum probability, which reported successful margin line generation for 7 cases out of 13 test cases. Table 4 shows that the average maximum distance between predicted and true margin lines was 194 μm, the average mean distance was 70.7 μm, and the average standard deviation of the distance was 45.6 μm (see Figure 6). Sample qualitative results in Figure 7 show minor differences between predicted and true margin lines.

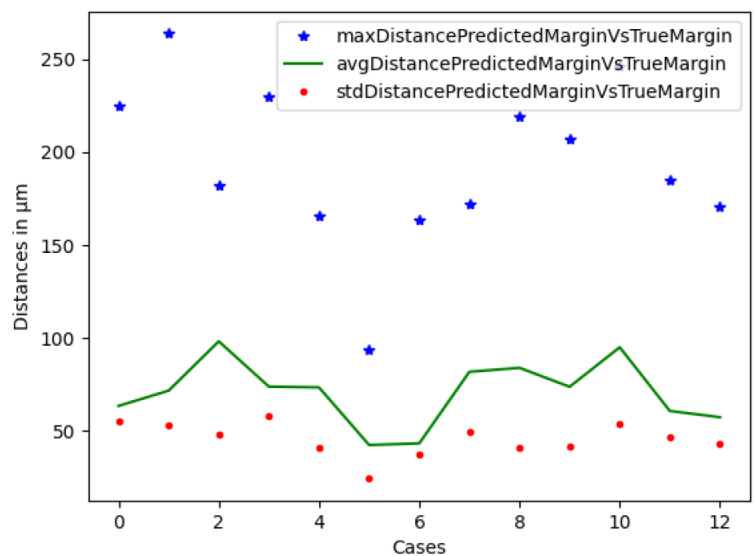

**Figure 6.** Maximum, average, and variances in the distances of predicted versus true margin line for 13 test cases.

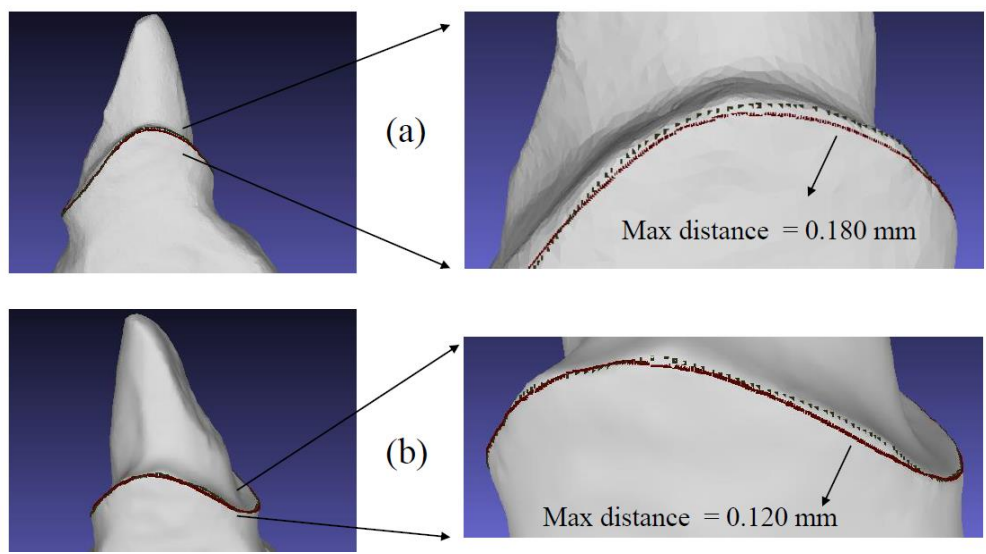

**Figure 7.** Qualitative results showing minor differences between predicted (red) and true margin lines (black) for test cases from positions 21 in (a) and 11 in (b).

### 3.2. Effect of decimation rate

Comparing model 1 with model 3 in Table 2, a decimation to 10 k triangles provided more successful cases (6 cases) compared with 20 k (4 cases). This is because a patch size of 9980 cells was fixed for both models. Therefore, a mesh with 20 k cells loses details about the margin line compared with a mesh with 10 k cells. In addition, the decimation algorithm favored the creation of cell boundaries along feature lines for the meshes decimated to 10 k cells.

### 3.3. Effect of input channels

The effect of input channels was investigated by performing an ablation study of the curvature channels (3 channels) as well as the vertex coordinate channels (9 channels). Comparing model 1 with model 2, adding the curvature increased the number of successful cases (Table 1). Comparing model 1 and model 4, we can see that removing the vertex positional channels decreased the number of successful cases by 3. The inclusion of curvature channels was essential as it reduced the fluctuations in the training and validation curves as shown in Figure 10 and it improved the margin line prediction results in terms of the number of successful test cases (Table 3).

### 3.4. Analysis of margin lines rating produced by dental technicians

The average margin line score rated by dental technicians was $2.65 \pm 0.66$ for the training data (41 samples) whereas it was $2.5 \pm 0.65$ for the test data (13 samples). Our data analysis revealed a correlation between the mean distance between true and predicted margin lines for test data and the rating of the true margin lines produced by dental technicians (Figure 8). The Spearman's rank correlation coefficient (r-value) was found to be -0.683, p-value = 0.010 (Table 4). In addition, we investigated the similarity of two margin lines, each produced by a different dental technician for the same dental die. By considering a set of 3 different cases, Figure 9 shows that the discrepancy between two margin lines produced for the same case could reach up to 200 μm.

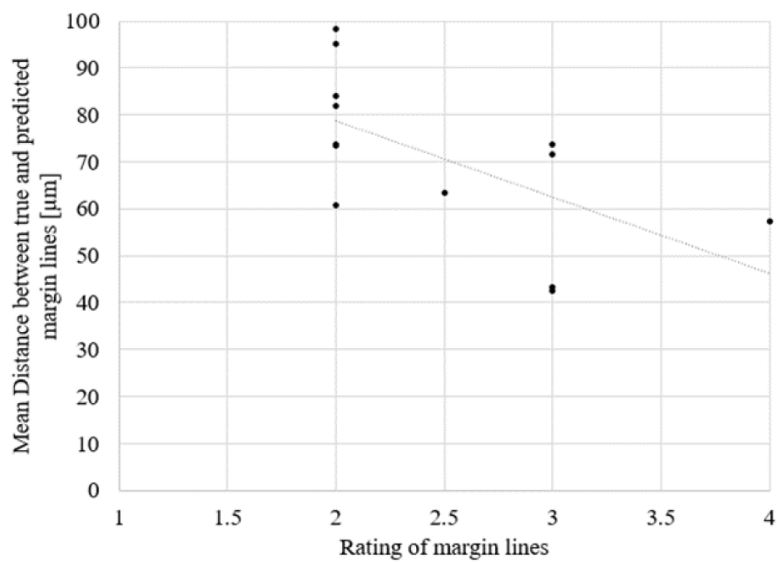

*Figure 8.* Correlation between the rating of true margin lines by dental technicians and the mean distance between true and predicted margin lines. Spearman's rank correlation coefficient = -0.683, p-value = 0.010.

**Table 2**
Results from models trained with different settings

| Model | Dice Similarity Coefficient (DSC) | Sensitivity (SEN) | Positive Predicted Values | No. successful cases |
|---|---|---|---|---|

| | DSC | SEN | PPV (PPV) | No. [out of 13] |
|---|---|---|---|---|
| 1 | 0.980±0.006 | 0.973±0.015 | 0.988±0.012 | 6 |
| 2 | 0.974±0.017 | 0.981±0.012 | 0.968±0.034 | 4 |
| 3 | 0.973±0.020 | 0.968±0.039 | 0.978±0.022 | 4 |
| 4 | 0.955±0.035 | 0.957±0.048 | 0.957±0.060 | 3 |

**Table 3**

Results of 5-folds of model 1 and comparison with ensemble model combined using two strategies: maximum probability and democracy

| Fold/metric | DSC (Accuracy) | SEN (Sensitivity) | PPV (Specificity) | No. successful cases [out of 13] |
|---|---|---|---|---|
| Fold 1 | 0.980±0.006 | 0.973±0.015 | 0.988±0.012 | 6 |
| Fold 2 | 0.971±0.022 | 0.976±0.016 | 0.967±0.047 | 5 |
| Fold 3 | 0.978±0.009 | 0.976±0.017 | 0.981±0.013 | 3 |
| Fold 4 | 0.974±0.018 | 0.969±0.014 | 0.979±0.032 | 4 |
| Fold 5 | 0.973±0.013 | 0.973±0.019 | 0.973±0.033 | 4 |
| Comb. max prob. | 0.981±0.006 | 0.974±0.015 | 0.989±0.011 | 7 |
| Comb. democracy | 0.981±0.006 | 0.977±0.012 | 0.986±0.015 | 6 |

**Table 4**

Ratings versus margin line prediction metrics for combined models with maximum probability.

| Prep ID | Rating (out of 4) | Max distance predicted vs. true margins [μm] | Mean distance predicted vs. true margins [μm] | Standard deviation predicted vs. true margins [μm] |
|---|---|---|---|---|
| 0862-21 | 2.5 | 225 | 63 | 55 |
| 6138-21 | 3 | 264 | 72 | 53 |
| 6158-11 | 2 | 182 | 98 | 48 |
| 6158-21 | 2 | 230 | 74 | 58 |
| 6223-21 | 2 | 166 | 73 | 41 |
| 6225-11 | 3 | 93 | 42 | 24 |
| 6225-21 | 3 | 163 | 43 | 37 |
| 6227-11 | 2 | 172 | 82 | 50 |
| 6227-21 | 2 | 219 | 84 | 41 |
| 6230-21 | 3 | 207 | 74 | 42 |
| 6231-21 | 2 | 246 | 95 | 54 |
| 6242-11 | 2 | 185 | 61 | 47 |
| 6260-11 | 4 | 171 | 57 | 43 |
| Average | 2.5 | 194 | 70.7 | 45.6 |

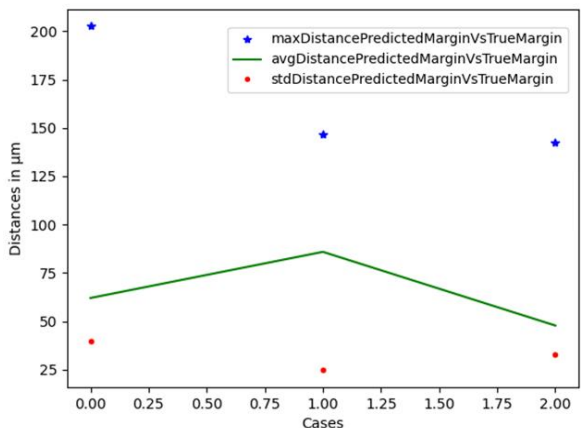
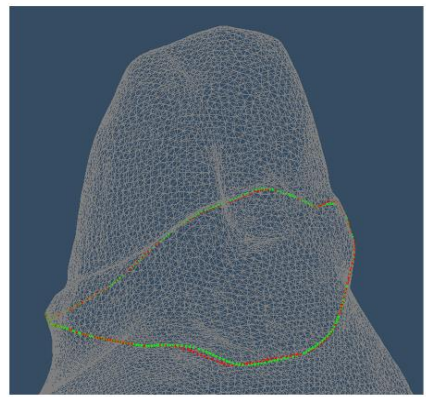

(a)　　　　　　　　　　(b)

**Figure 9.** (a) Maximum and average distances between two margin lines produced by different technicians in blue and red, respectively. The average maximum distance is 164 μm and the average mean distance is 65.2 μm. (b) Qualitative comparison between two margin lines produced by two different technicians for the same case.

The inclusion of coordinates and curvature channels was essential to improve the accuracy of the model and the quality of margin line prediction results (Table 3). Because the curvature features were reported to be essential for margin lines, curvature features were added as input channels to train the neural network. Figure 10 shows that curvature channels reduced the fluctuations in the training and validation curves and reduced the validation loss improving the accuracy of the trained AI model.

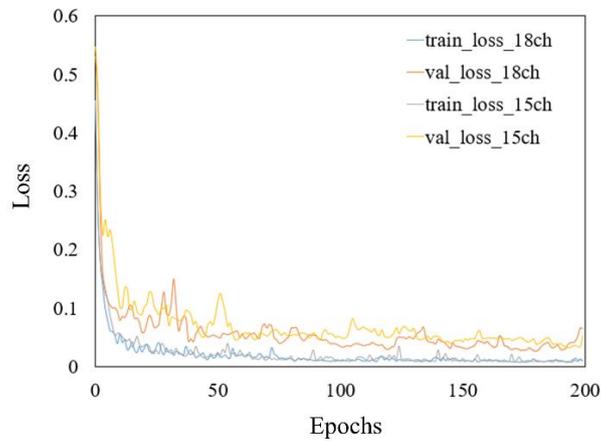

**Figure 10.** Comparison between training and validation loss curves for the segmentation model trained with (18 channels) and without (15 channels) curvature channels.

*3.5. Analysis of unsuccessful test cases*

In this section, we discuss the features of unsuccessful test cases. Figure 11 shows the predicted segmentation of crown bottoms of 4 test cases along with the ground truth (green) and predicted (red) margin lines. The figure shows that the predicted margin line for case (a) is slightly above the correct position. This case possesses a pump at the margin line location which confuses the algorithm. A similar situation applies for case (b). For case (c), the figure shows that the margin line is slightly under the correct position on the front side and slightly above the correct position on the back side due to the presence of pumps on both sides. In case (d), the algorithm predicted the margin line below the correct position due to an error from the segmentation algorithm. The deep learning model was confused during die segmentation due to the presence of multiple nearby features with high curvature.

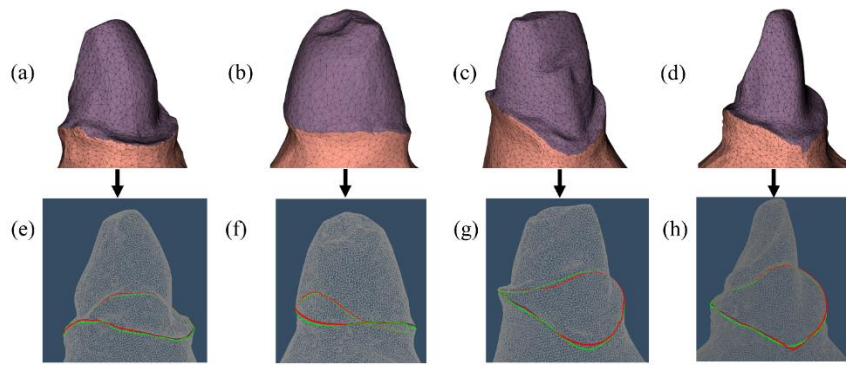

**Figure 11.** Predicted segmentation of crown bottoms of 4 test cases along with the ground truth (green) and predicted (red) margin lines.

## 4. Discussion

Advances in digital dentistry and artificial intelligence have increased the demands for automatic digital design and manufacturing. Among recent research advances in the field is a robot-assisted tooth preparation based on augmented reality [29] and automatic crown generation using deep learning [14]. Hence, the need for automatic and accurate margin line extraction techniques is crucial. The use of AI-assisted margin line extraction tools promises a high potential to solve the margin line generation problem [29]. However, it is difficult to generalize about the level of accuracy of AI for the detection of tooth preparation margins due to the small number of AI-based systems specified for margin line extraction (just one study to date). As a result, more research is needed in this area [29]. While it is essential for a margin line to be the outermost region, this does not necessarily always mean the region with the highest curvature. This makes the process of automatically and accurately extracting an accurate margin challenging.

The MeshSegNet architecture was chosen from a wide range of networks available in the literature for point cloud segmentation [30] because it proved high accuracy on segmentation of dental data and identification of dental features. Although the architecture of MeshSegNet was used, the novelty of this work lies in re-training the network with different input channels to detect margin lines on the surfaces of prepared teeth. Our application entails segmenting the tooth itself into two regions then extracting a margin line through spline fitting, rather than segmenting a particular tooth from a scanned jaw. Since curvature features were reported to be essential for margin lines [31], curvature features were added as input channels to train the neural network. Our work also reveals that using an ensemble technique on MeshSegNet trained models helps increase the accuracy of the predicted margin lines. In addition, we propose for the first time a novel technique to generate accurate ground truth labels for margin lines based on actual physical crown designs produced by dental professionals. We also presented, for the first time, a unique analysis to correlate the quality of preparations produced by dentists with the quality of the margin lines predicted by the proposed framework. The boundary faces of the resulting segmented mesh (in particular the segmented crown bottom mesh) was used to create the spline representing the predicted margin line. A spline was approximated to best fit the centers of the boundary faces of the segmented mesh to predict the margin line. The Euclidean distance metric was used to estimate the total error between the ground truth margin lines and the predicted splines. In order to do that, margin lines were discretized to 5,000 points each on the splines. After that the Euclidean distance between every two closest points on the predicted and ground truth margin lines was calculated. Table 4 in the shows the maximum of that Euclidean distance for each test case and also shows the average maximum distance for all the test cases to be 194 μm. Out of 13 unseen test cases, the smallest maximum distance was 93 μm and the greatest was 264 μm.

The clinically accepted accuracy for the margin line region is 100 μm which corresponds to the accuracy of an intraoral scanner (IOS) [32]. On the other hand, we found that discrepancies in distances between two real margin lines produced by two different technicians for the same case could reach up to 200 μm (Figure 9). The best results obtained with our combined model with maximum probability showed an average maximum distance of 194 μm between predicted and true margin lines, which is considered very close to the human error reported in Figure 9. The 4th column of Table 4 considers the mean Euclidean distance between the ground truth and predicted margin lines and reports it for every test case. The average of this mean Euclidean distance was found to be 70.7 μm. The reported results do not violate the distance threshold of 200 μm representing human error. These results were achieved with the combined model with maximum probability, which also showed an average standard deviation of 45.6 μm between the predicted and true margin lines. This result is within the range of accepted accuracy and was achieved with a test set that constituted about 25% of the total dataset. Compared with [8], our proposed model achieves a higher accuracy (DSC = 0.981±0.006) with less data. Unfortunately, there are no other studies in literature dealing with margin line detection on tooth preparations using AI [29]. In addition, it is illogical to compare our results with the only available study in the literature [8], which claimed up to 97% accuracy, because the data are not the same. In fact, dental data is highly complex and variable such that the presence of a couple of difficult cases could cause substantial changes in the accuracy score.

Generally, due to the high localization of the margin line as well as the highly required accuracy and precision, small differences in the segmentation metric results lead to large variances in the relative distances between predicted and true margin lines. Therefore, the sensitivity metric is extremely important in that regard. In fact, with a higher sensitivity score, a model will more likely predict a margin line on or below the true margin line which is more practical than a prediction above the true margin line. In other words, it is extremely important to minimize the false negatives (see Figure 5). Combining the 5 folds with maximum probability led to the maximum number of successful test cases (Table 3). In 7 out of 13 test cases, the predicted margin lines by the ensemble model with maximum probability had a maximum distance smaller than or equal to 200 μm.

The average rating of true margin line was reported by the technicians to be 2.5 out of 4. This indicates the presence of a variant and problematic margin line in some cases and highlights the importance of automatic and consistent margin line extraction. Since the absolute r-value of the correlation was greater than 0.5, a relationship exists between the ratings of dental technicians for the true margin lines and the model predictions of the mean distance between the predicted and true margin lines. A negative r-value of -0.683 indicates that as the rating of tooth preparation increases, the difference between the predicted margin and true margin lines decreases. This also indicates that our model naturally produces better margin lines as the quality of dental preparation increases.

The automatic generation of margin lines could lead to automated teeth preparation for crown treatment with help of simulated reality and digital cameras and even replace the freehand method of creating a preparation [17]. Although the accuracy achieved by the proposed model is acceptable, there is still room for improvement. It is expected that increasing the dataset size will increase the prediction accuracy.

The size of our dataset was limited by the availability of data. We attempted to mitigate this issue by: (1) using various well-known augmentation techniques to extensively increase the number of our training dataset (each sample was augmented 20 times); (2) focusing our study on a particular type of teeth to reduce the variability given the limited amount of data (only 4 positions out of 28 positions were considered, which are 11, 21, 31, and 41); (3) registering the data by rotating 180° the labeled dies belonging to upper arches such that they share the same orientation with dies belonging to lower arches further reduced the 4 positions to only two positions that are also symmetric. For example: positions 11, 21, 13, 14 would be more different than 11, 21 ,31, 41 (which are the positions used in our experiments).

Although the proposed framework was trained and tested on incisor teeth, it can be considered a novel first step towards a fully automated and accurate margin line detection and could be easily extended to other types of teeth. Obviously, increasing the dataset increases the deep learning model accuracy. Nevertheless, this work contributes a novel end-to-end framework for margin line extraction on prepared teeth more than a ready-to-use pretrained model. The next step is to scale-up the model training on a larger dataset for all teeth positions and deploy a generic model online for external users. We aim to provide software to help dental technicians produce better margin lines during the crown generation process. The software interface will also be simple enough to be easily used and manipulated by interested dentists, dental students, and dental technologists. Future development will also focus on enhancing the preprocessing steps to improve the triangulation resolution of tooth meshes, particularly near the margin line. Adaptive mesh refinement will be introduced based on curvatures such as regions of high curvature will be more decimated. Although this enhancement increases the number of points available for more accurate spline prediction of the ground truth margin, it also results in greater preprocessing time and memory usage. As such, a balance must be achieved between accuracy and computational efficiency. Another future direction is to improve the postprocessing pipeline to ensure the extraction of a margin line that avoids undercuts.

## 5. Conclusion

This paper presented a data-efficient deep learning framework for margin line extraction from dental preparations. The successful identification of a proper margin line is a key factor in the success of the dental crown design. The proposed end-to-end framework extracted margin lines with acceptable relative distances from the true margin lines designed by dental technicians for incisor teeth. The average maximum distance obtained with an ensemble model combined with maximum probability was 194 µm, which was close to the human measured error of 200 µm. It was also demonstrated that the better the quality of the preparation, the better the prediction of the margin line by the proposed model.

We made the training and test datasets available to enable the community to compare other models with the model used in this work for the same dataset[2]. Among the limitations of our study are the small number of data used in our experimental set up and the restriction of the teeth type to the first incisor. The future directions involve scaling up the dataset with more tooth types including other canines, incisors, premolars, and molars from both the upper and lower jaws and increasing the total number of training and test data. Future development will also entail improving the preprocessing procedure to increase the triangulation resolution of the teeth meshes near the margin line region and improving the postprocessing by smart correction of segmentation results to extract more accurate margin lines.

---

[2] https://github.com/intellident-ai/teethPreparationData/tree/main

**Acknowledgments:** Author Francois Guibault acknowledges the funding received by the Natural Science and Engineering Research Council of Canada [Ref: ALLRP 549122-2019], Institut de valorisation des données (IVADO) [Ref: PostDoc-2020a-5943530233], and MEDTEQ [19-D Volumétrie dentaire 2]. Author Ammar Alsheghri acknowledges the funding received from KFUPM [Ref: EC241009]. We thank the Canadian Digital Alliance for providing the computational resources used in this work. The authors acknowledge the help and support from JACOBB and Comet Technologies Inc. (Formally Object Research Systems Inc). This work has not been submitted for publication or presentation elsewhere. Author Ammar Alsheghri acknowledges the guidance of his mentor at KFUPM, Dr. Muhammad Hawwa.

**Appendix:**

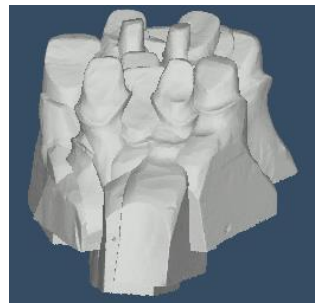

**Figure A1.** Dies sharing similar orientations as a preprocessing step before training.

**References**

1. Podhorsky, A.; Rehmann, P.; Wöstmann, B. Tooth preparation for full-coverage restorations—a literature review. *J Clinical oral investigations*, 2015. **19**(5): p. 959-968.
2. Ramesh, G.; Nayar, S.; Chandrakala, S. Principles of Tooth Preparation-Review Article. *J Indian Journal of Forensic Medicine Toxicology*, 2020. **14**(4).
3. Yu, N., Dai, H.W., Tan, F.B., Song, J.L., Ma, C.Y. and Tong, X.L. Effect of different tooth preparation designs on the marginal and internal fit discrepancies of cobalt-chromium crowns produced by computer-aided designing and selective laser melting processes. *The Journal of Advanced Prosthodontics*, 2021. **13**(5): p. 333.
4. Seymour, K., D. Samarawickrama, and E. Lynch, Metal ceramic crowns--a review of tooth preparation. *The European journal of prosthodontics restorative dentistry*, 1999. **7**(2): p. 79-84.
5. Abdullah, A.O., Muhammed, F.K., Zheng, B. and Liu, Y. An overview of computer aided design/computer aided manufacturing (CAD/CAM) in restorative dentistry. *Journal of Dental Materials Techniques*, 2018. **7**(1): p. 1-10.
6. Yu, H., Chen, Y.H., Cheng, H. and Sawase, T. Finish-line designs for ceramic crowns: a systematic review and meta-analysis. *J The Journal of Prosthetic Dentistry*, 2019. **122**(1): p. 22-30.
7. Reich, S., A. Petschelt, and U. Lohbauer, The effect of finish line preparation and layer thickness on the failure load and fractography of ZrO2 copings. *J The Journal of prosthetic dentistry*, 2008. **99**(5): p. 369-376.
8. Zhang, B., Dai, N., Tian, S., Yuan, F. and Yu, Q. The extraction method of tooth preparation margin line based on S-Octree CNN. *International Journal for Numerical Methods in Biomedical Engineering*, 2019. **35**(10): p. 3241.
9. Ni, H., Lin, X., Ning, X. and Zhang, J, Edge detection and feature line tracing in 3D-point clouds by analyzing geometric properties of neighborhoods. *Remote Sensing*, 2016. **8**(9): p. 2072-4292.
10. ZHANG, C., Extraction of Dental Biological Feature Line Based on Heuristic Search Strategy. *China Mechanical Engineering*, 2012. **23**(13): p. 1567.
11. Li, X., Wang, X. and Chen, M. Accurate extraction of outermost biological characteristic curves in tooth preparations with fuzzy regions. *J Computers in Biology Medicine*, 2018. **103**: p. 208-219.
12. Shan, T., Tay, F.R. and Gu, L. Applications of artificial intelligence in dentistry: A comprehensive review. *Journal of Esthetic and Restorative Dentistry* 34.1 (2022): 259-280.
13. Alshadidi, A.A.F., Alshahrani, A.A., Aldosari, L.I.N., Chaturvedi, S., Saini, R.S., Hassan, S.A.B., Cicciù, M. and Minervini, G. Investigation on the application of artificial intelligence in prosthodontics. *Applied Sciences* 13.8 (2023): 5004.

14. Lessard, O., Guibault, F., Keren, J. and Cheriet, F. Dental Restoration using a Multi-Resolution Deep Learning Approach. In Proceedings of IEEE 19th International Symposium on Biomedical Imaging (ISBI), Kolkata, India, 28-31 March 2022; IEEE: New York City, United States, 26 April 2022.

15. Hosseinimanesh, G., Ghadiri, F., Alsheghri, A., Zhang, Y., Keren, J., Cheriet, F. and Guibault, F. Improving the quality of dental crown using a transformer-based method. In Medical Imaging 2023: Physics of Medical Imaging (Vol. 12463, pp. 802-809). SPIE.

16. Piché, N., et al,. (2023). U.S. Patent Application No. 18/017,809, filed 2023 Sep 7.

17. Tian, S., Wang, M., Ma, H., Huang, P., Dai, N., Sun, Y. and Meng, J. Efficient tooth gingival margin line reconstruction via adversarial learning. *J Biomedical Signal Processing Control*, 2022. **78**: p. 103954.

18. Jiang, X., Xu, B., Wei, M., Wu, K., Yang, S., Qian, L., Liu, N. and Peng, Q. C2F-3DToothSeg: Coarse-to-fine 3D tooth segmentation via intuitive single clicks. *J Computers Graphics*, 2022. **102**: p. 601-609.

19. Alsheghri, A., Ghadiri, F., Zhang, Y., Lessard, O., Keren, J., Cheriet, F. and Guibault, F. Semi-supervised segmentation of tooth from 3D scanned dental arches. in Medical Imaging 2022: Image Processing (Vol. 12032, pp. 766-771). SPIE.

20. Lian, C.; Wang, L.; Wu, T.H.; Wang, F.; Yap, P.T.; Ko, C.C.; Shen, D. Deep multi-scale mesh feature learning for automated labeling of raw dental surfaces from 3D intraoral scanners. *IEEE transactions on medical imaging* 2020. *39(7)*, p. 2440-2450.

21. Im, J., Kim, J.Y., Yu, H.S., Lee, K.J., Choi, S.H., Kim, J.H., Ahn, H.K. and Cha, J.Y. "Accuracy and efficiency of automatic tooth segmentation in digital dental models using deep learning." *Scientific reports* 12.1 (2022): 9429.

22. Tabatabaian, F., Vora, S. R., Mirabbasi, S. Applications, functions, and accuracy of artificial intelligence in restorative dentistry: A literature review. *Journal of Esthetic and Restorative Dentistry*, 2023, 35(6), 842-859.

23. Qi, C.R., Su, H., Mo, K. and Guibas, L.J. Pointnet: Deep learning on point sets for 3d classification and segmentation. in Proceedings of the IEEE conference on computer vision and pattern recognition. 2017.

24. Zhang, R., Li, G., Wunderlich, T. and Wang, L. A survey on deep learning-based precise boundary recovery of semantic segmentation for images and point clouds. *International Journal of Applied Earth Observation Geoinformation*, 2021. **102**: p. 102411.

25. Rodrigues, R.S., J.F. Morgado, and A.J. Gomes. Part-based mesh segmentation: a survey. in *Computer Graphics Forum*. 2018. Wiley Online Library.

26. Taha, A.A.; Hanbury, A. Metrics for evaluating 3D medical image segmentation: analysis, selection, and tool. *J BMC medical imaging*, 2015. **15**(1): p. 1-28.

27. Goodacre, C.J., Campagni, W.V. and Aquilino, S.A. Aquilino. "Tooth preparations for complete crowns: an art form based on scientific principles." *The Journal of prosthetic dentistry* 85.4 (2001): 363-376.

28. Blair, F.M., Wassell, R.W. and Steele, J.G. "Crowns and other extra-coronal restorations: Preparations for full veneer crowns." British dental journal 192.10 (2002): 561-571.

29. Jiang, J., Guo, Y., Huang, Z., Zhang, Y., Wu, D., Liu, Y. Adjacent surface trajectory planning of robot-assisted tooth preparation based on augmented reality. Engineering Science Technology, an International Journal, 2022. **27**: p. 101001.

30. Xie, Y., Tian, J., Zhu, X.X. Linking points with labels in 3D: A review of point cloud semantic segmentation. J IEEE Geoscience Remote Sensing Magazine, 2020. **8**(4): p. 38-59.

31. Kuralt, M., Cmok Kučič, A., Gašperšič, R., Grošelj, J., Knez, M., Fidler, A. Gingival shape analysis using surface curvature estimation of the intraoral scans. BMC Oral Health, 2022. **22**(1): p. 1-11.

32. Son, K.; Lee, K. B. Effect of finish line locations of tooth preparation on the accuracy of intraoral scanners. *J Int. J. Comput. Dent*, 2021. **24**: p. 29-40.

33. Dierckx, P. Curve and surface fitting with splines, *Monographs on Numerical Analysis*, 1993, Oxford University Press.